\documentclass[aps,twocolumn,preprintnumbers,amsmath,amssymb,longbibliography]{revtex4-1}

\usepackage{graphicx}
\usepackage{dcolumn}
\usepackage{bm}
\usepackage{multirow}
\usepackage{color}
\usepackage[normalem]{ulem}
\usepackage{amsmath}
\usepackage{gensymb}
\usepackage[colorlinks=true]{hyperref}

\definecolor{mblue}{rgb}{0,0.35,0.75}

\begin{document}
\title{Exciton states in monolayer MoSe$_2$ and MoTe$_2$ probed by upconversion spectroscopy}

\author{B. Han$^1$}
\author{C. Robert$^1$}
\author{E.~Courtade$^1$}
\author{M. Manca$^1$}
\author{S. Shree$^1$}
\author{T. Amand$^1$}
\author{P. Renucci$^1$}
\author{T.~Taniguchi$^2$}
\author{K. Watanabe$^2$}
\author{X. Marie$^1$}
\author{L. E. Golub$^3$}
\author{M. M. Glazov$^3$}
\email{glazov@coherent.ioffe.ru}
\author{B. Urbaszek$^1$}
\email{urbaszek@insa-toulouse.fr}

\affiliation{%
$^1$Universit\'e de Toulouse, INSA-CNRS-UPS, LPCNO, 135 Av. Rangueil, 31077 Toulouse, France}
\affiliation{$^2$National Institute for Materials Science, Tsukuba, Ibaraki 305-0044, Japan}
\affiliation{$^3$Ioffe Institute, 194021 St.\,Petersburg, Russia}

\begin{abstract}
Transitions metal dichalcogenides (TMDs) are direct semiconductors in the atomic monolayer (ML) limit with fascinating optical and spin-valley properties. The strong optical absorption of up to 20 \% for a single ML is governed by excitons, electron-hole pairs bound by Coulomb attraction. 
Excited exciton states in MoSe$_2$ and MoTe$_2$ monolayers have so far been elusive due to their
low oscillator strength and strong inhomogeneous broadening. Here we show that encapsulation in hexagonal boron nitride results in emission line width of the A:1$s$ exciton below 1.5~meV and 3~meV in our MoSe$_2$ and MoTe$_2$ monolayer samples, respectively. This allows us to investigate the excited exciton states by photoluminescence upconversion spectroscopy for both monolayer materials. The excitation laser is tuned into resonance with the A:1$s$ transition and we observe emission of excited exciton states up to 200~meV above the laser energy. We demonstrate bias control of the efficiency of this non-linear optical process. 
At the origin of upconversion our model calculations suggest an exciton-exciton (Auger) scattering mechanism specific to TMD MLs involving an excited conduction band thus generating high energy excitons with small wave-vectors. The optical transitions are further investigated by white light reflectivity, photoluminescence excitation and resonant Raman scattering confirming their origin as excited excitonic states in monolayer thin semiconductors.
\end{abstract}

\maketitle
\section{Introduction}

Transition metal dichalcogenides such as MoS$_2$, WS$_2$, WSe$_2$, MoSe$_2$ and MoTe$_2$ are direct band gap semiconductors when thinned down to one monolayer \cite{Novoselov:2016a,Geim:2013a,Mak:2010a, Splendiani:2010a, Wang:2012c,Mak:2016a}. Their bandgap is situated in the visible to near infrared of the optical spectrum. Since the Coulomb interaction is strong in this ultimate 2D limit, the optical properties are dominated by excitons, bound electron-hole pairs~ \cite{Wang:2017b,He:2014a,Ugeda:2014a,Chernikov:2014a,Ye:2014a,Qiu:2013a,Ramasubramaniam:2012a,Wang:2015b}. 
Recently encapsulation in hexagonal boron nitride (hBN) of TMD monolayers (MLs) has resulted in considerable narrowing of the exciton transition linewidth down to 1~meV \cite{Jin:2016a,Manca:2017a,Chow:2017a,Cadiz:2017a,Ajayi:2017a,Wang:2016v,wierzbowski2017direct}. This gives now access to fine features of the exciton spectra that dominate the linear and non-linear optical properties. 

\begin{figure}[t]
\includegraphics[width=0.47\textwidth,keepaspectratio=true]{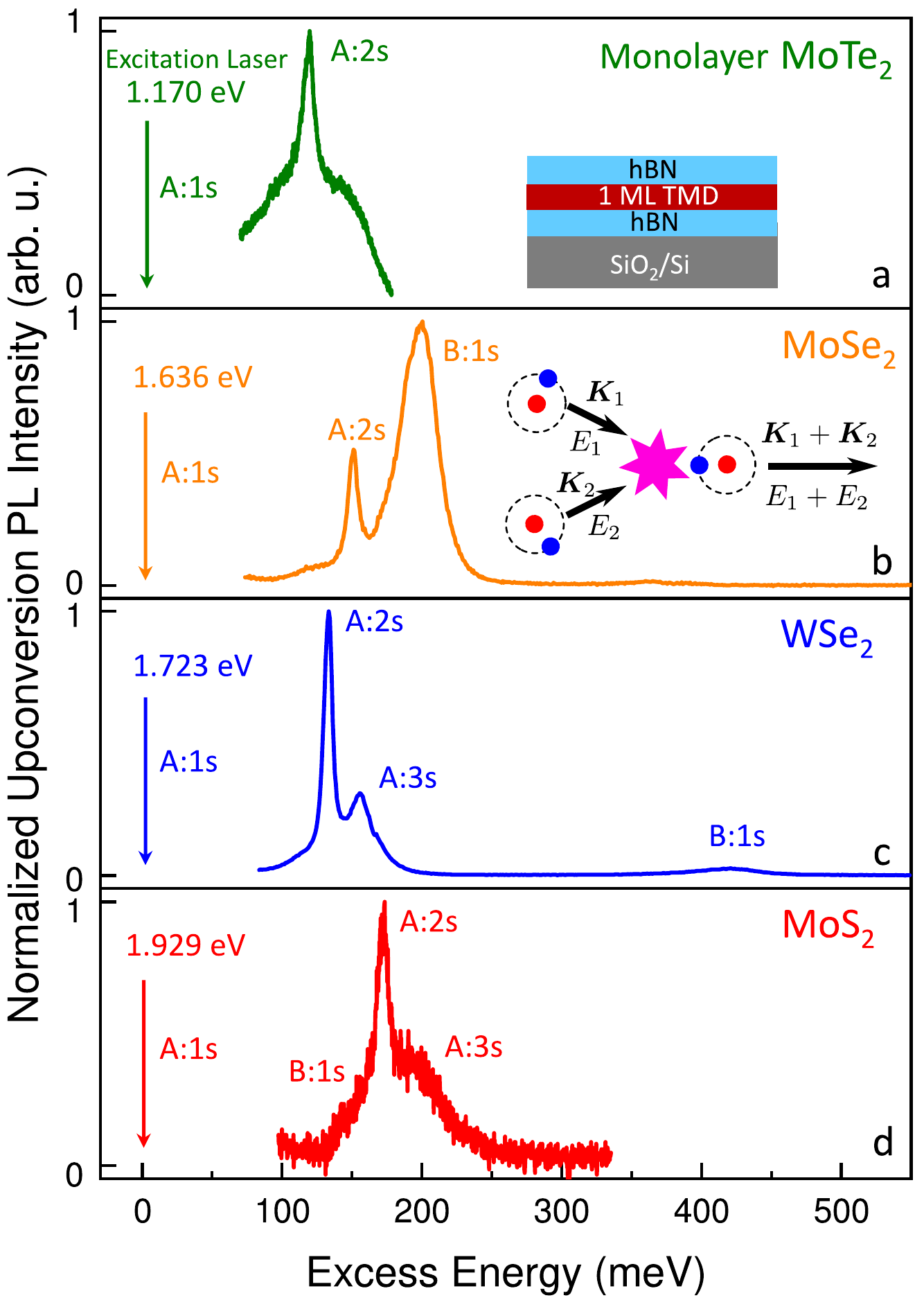}
\caption{\textbf{Upconversion spectroscopy in TMD monolayers.} We present for four different monolayer materials resonant excitation experiments of the A:$1s$ exciton at $T=4$~K, that result in PL emission at higher energy. The laser energy -- equal to the A:$1s$ transition energy -- is marked by a vertical arrow. The upconversion emission peaks are labeled A:$2s$ and B:$1s$, where the origin of these peaks is confirmed in complementary experiments such as reflectivity and PLE.
The results for WSe$_2$ are reproduced from \cite{Manca:2017a}, the MoS$_2$ results from \cite{Robert:2018a}. Inset in (a) shows a scheme of the sample. Inset in (b) shows the exciton-exciton Auger process where one exciton annihilates and another one acquires total momentum and energy of the two particles.}
\label{fig:fig1} 
\end{figure}

Optical excitation of a semiconductor at the bandgap typically results in luminescence at lower energy due to energy relaxation of charge carriers and excitons. In Fig.~\ref{fig:fig1}a-d we show that resonant laser excitation of the lowest energy exciton resonance A:1$s$ results in pronounced photoluminescence (PL) emission at \textit{higher} energy than the excitation laser for four different TMD ML materials. This effect is generally termed upconversion and has been observed for different semiconductor structures such as InP/InAs heterojunctions, CdTe quantum wells and InAs quantum dots \cite{Seidel:1994a,Hellmann:1995a,Poles:1999a,Paskov:2000a,Chen:2012a} albeit based on different microscopic mechanisms. Upconversion has  previously been reported for WSe$_2$ \cite{Manca:2017a,Jones:2015a} and MoS$_2$ \cite{Robert:2018a} MLs. 
These experiments allow detailed insight into the light matter interaction physics of excitons in TMD monolayers: First, clarifying the origin of upconversion signal is in itself a crucial problem, as the origin of excess energy needs to be identified and the role of exciton-exciton scattering mechanisms is revealed.
Second, upconversion spectroscopy gives us access to the excited exciton states that govern absorption and emission, so far not well understood in ML MoSe$_2$ and MoTe$_2$. The centre of mass motion of excitons, in analogy to the hydrogen atom and positronium, is characterized by a principle quantum number $n=1,2,3….$~, where typical photoluminescence emission stems from the $n=1$ exciton of the A-exciton series, labelled A:1$s$. The optical absorption in energy above the A:1$s$ optical transition will be determined by the excited states A:2$s,3s$ etc and the B-exciton series, separated from the A-exciton mainly by the spin-orbit splitting in the 200-400~meV range \cite{Kormanyos:2015a}. 
We show that upconversion allows us to access excited exciton states for MoSe$_2$ and MoTe$_2$, which is not possible in samples that are not encapsulated in hBN as the excited A-excitons spectrally overlap with the B-exciton series. We demonstrate bias control of the upconversion process. We provide an in-depth study of exciton states in MoTe$_2$ comparing upconversion with photoluminescence excitation spectroscopy (PLE) and white light reflectivity. In the last part of the paper we provide a theoretical model and discuss the origin of upconversion in TMD monolayers. Our model calculations suggest that Auger type exciton-exciton scattering is very efficient in TMD MLs as compared to other semiconductor nano-structures due to (i) the strong Coulomb interaction, which makes it possible to relax the single-electron momentum conservation~\cite{PhysRevB.54.16625,PhysRevB.73.245424} and (ii) the possibility of a resonant processes involving exciton transfer to an excited energy band. 

\section{Excited state spectroscopy in ML $\mbox{MoSe}_2$}
\label{sec:mose2}

\begin{figure*}
\includegraphics[width=0.70\textwidth,keepaspectratio=true]{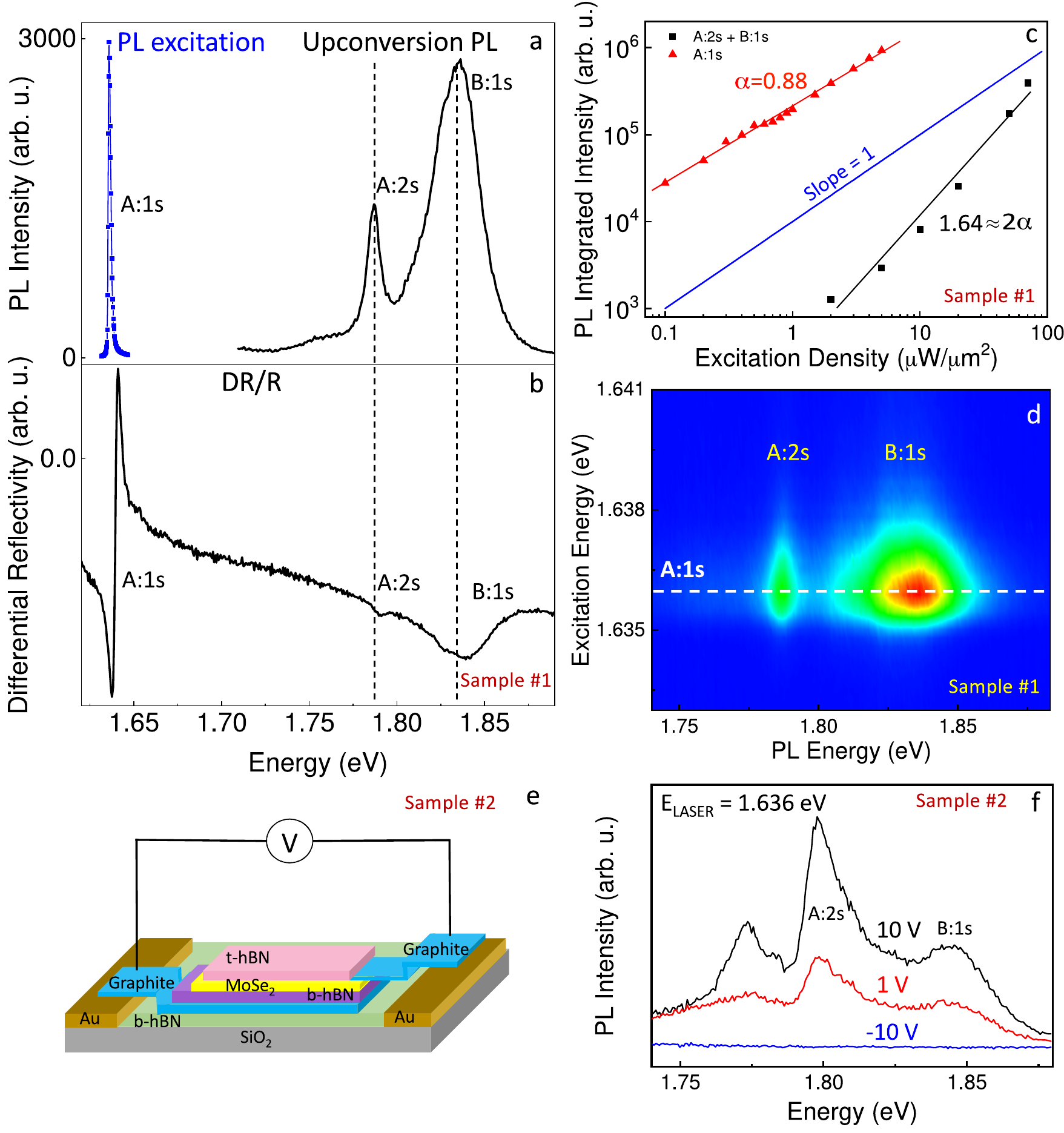}
\caption{\textbf{Control of upconversion in ML MoSe$_2$.} T$=4$~K. Sample \#1(a): Scanning a Ti-Sa laser across the A:$1s$ transition results in upconversion emission of the A:$2s$ and B:$1s$ transitions (black curve for excitation exactly at resonance). Blue symbols give the integrated upconversion intensity as a function of laser energy. (b) Reflection contrast for the same sample spot, confirming the energy positions of A:1$s$, A:2$s$ and B:1$s$ transtions. (c) The power dependence of the upconversion signal shows an increase with a slope $\alpha$ of roughly 1.64 (black symbols), as compared for the standard A:$1s$ exciton emission with a slope roughly half (0.88 - red symbols). (d) Contour plot (blue - below 50 counts; red $>2000$ counts) of upconversion PL intensity as the excitation laser is swept across the A:$1s$ resonance.  Sample \#2: (e) Schematics of the charge tunable device. (f) Voltage control of upconversion. The signal is maximal in the neutral regime and gets weaker as the $n$-type regime favours trion and not neutral exciton absorption, the B:1$s$ and A:2$s$ emission are marked, a third emission peak of yet to be determined origin appears at lower energy. }
\label{fig:fig2} 
\end{figure*}

Monolayer MoSe$_2$ is a very versatile TMD material ideally suited to explore coupling to optical cavities \cite{dufferwiel2017valley,lundt2016monolayered}, investigating voltage control of monolayer mirrors \cite{Scuri:2018a,back2017giant} and interplay between charged and neutral excitons \cite{hao2016coherent}. Most of these experiments are based on the optical response of the lowest energy exciton state A:1$s$, but very little is known about excited exciton states that govern optical absorption at higher energies and energy relaxation pathways for PL emission. \\
\indent The experimental results for the high quality MoSe$_2$ samples encapsulated in hBN \cite{Taniguchi:2007a} are summarized in Fig.~\ref{fig:fig2}, details of the experimental set-up can be found in Appendix~\ref{app:A}. In differential white light reflectivity at T$=4$~K, we clearly observe the A and B-exciton $1s$ states \cite{Ross:2013a,Wang:2015a,Wang:2015c}, the A:$1s$ resonance has a full width at half maximum (FWHM) of the order of 2~meV \cite{shree2018exciton}. 
In Fig.~\ref{fig:fig2}a we show an intriguing result: excitation of the sample with a low power, continues wave (cw), narrow linewidth ($<1~\mu$eV) laser at the A:$1s$ energy results in emission of the B:$1s$ transition at \textit{higher} energy. As we scan the laser across the A:1$s$ resonance, the PL intensity (black graph) has a clear maximum in intensity when the laser is exaclty at the A:1$s$ resonance, the blue data points represent the integrated upconversion intensity for different laser energies. An additional transition to B:1$s$ appears about 150~meV above the A:$1s$ in upconversion PL that we tentatively assign to the excited A-exciton, A:2$s$ state. This transition is also visible in reflectivity in Fig.~\ref{fig:fig2}b. For samples directly exfoliated onto SiO$_2$ the excited A-exciton states were not directly accessible due to their overlap with the B-exciton $1s$ state. A fingerprint of the A:$2p$ state was reported in two-photon PL excitation (PLE) experiments~\cite{Wang:2015c}, where the B:$1s$ state absorption is strongly reduced~\cite{Glazov:2017a}.\\
\indent  Now we investigate the origin of the upconversion PL in ML MoSe$_2$ shown in Figs.~\ref{fig:fig1}b and \ref{fig:fig2}a. We compare the evolution of standard and upconversion PL intensity as a function of laser power in Fig.~\ref{fig:fig2}c . The slope of the upconversion intensity versus laser power ($1.64$) is roughly twice as high as for standard PL ($0.88$), consistent with a two-photon (two exciton) process being at the origin of this non-linear optical effect. In Fig.~\ref{fig:fig2}d we plot the upconversion emission as the laser is scanned across the A:1$s$ resonance. Upconversion is only detectable over a 2~meV range of laser energy when the laser is in resonance with the A:$1s$ state. This indicates upconversion is a resonant process, as observed for WSe$_2$ monolayers \cite{Manca:2017a}. This conclusion gets additional support from upconversion experiments in a charge tunable device (Fig.~\ref{fig:fig2}e), presented in Fig.~\ref{fig:fig2}f. At a bias of +10~V, the excitation laser is tuned into resonance with the neutral A:$1s$ state. As the applied voltage is lowered to +1~V, electrons are added to the monolayer, decreasing absorption strength at the neutral exciton resonance. The upconversion signal is not detectable any more for a bias of $-10$~V as absorption at the exciton resonance is inefficient, as the charged exciton absorption, at a different energy, dominates \cite{Scuri:2018a,back2017giant,Ross:2013a}. Differential reflectivity of our device shows a strong neutral exciton resonance at +10~V whereas for $-10$~V the charged exciton transition dominates, see Reflectivity measurements on the device in Appendix \ref{app:data}. These experiments confirm that upconversion PL emission has its origin in resonant neutral exciton generation and can be controlled electrically in charge tunable structures.
\begin{figure*}
\includegraphics[width=0.7\textwidth,keepaspectratio=true]{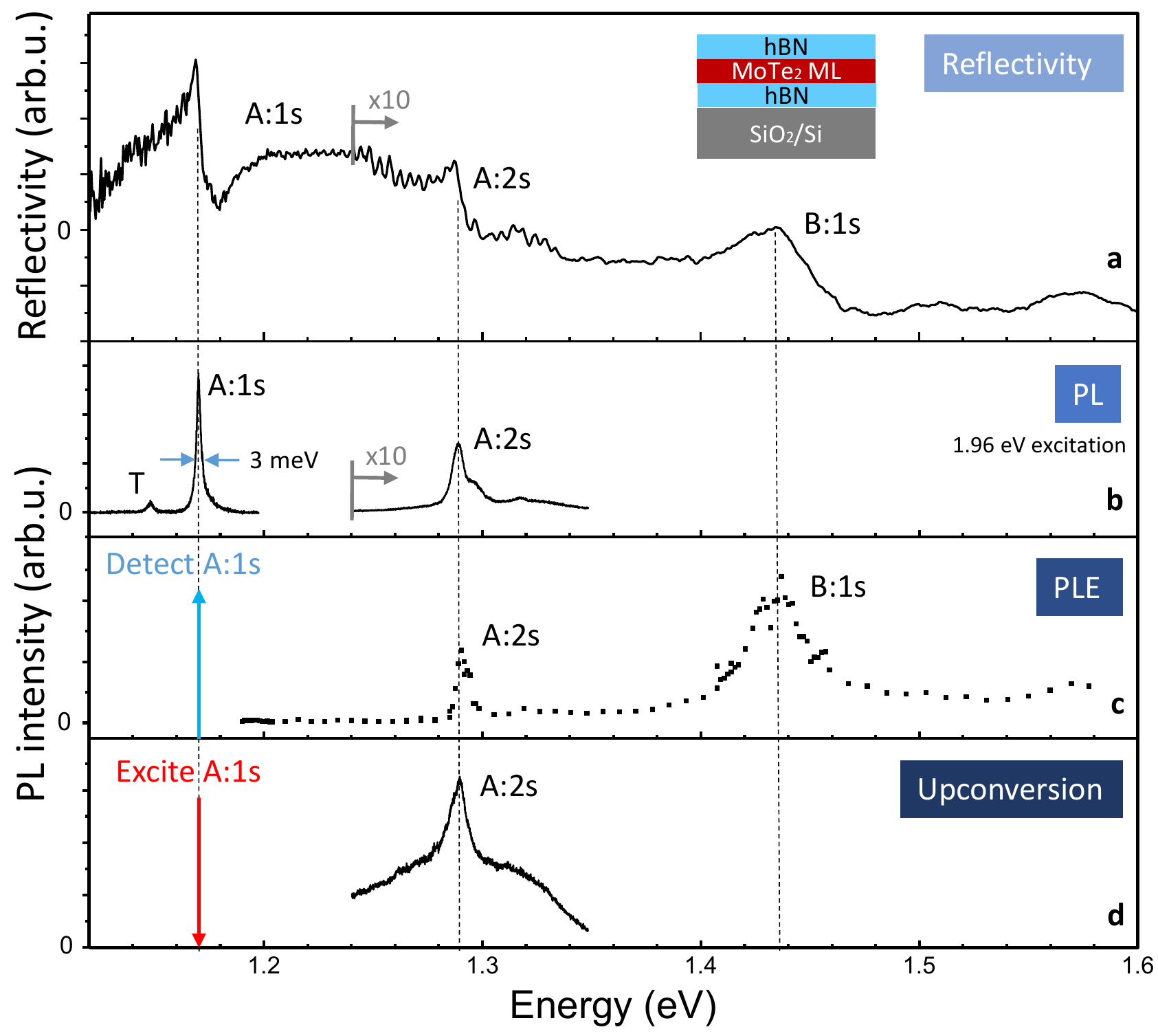}
\caption{\textbf{Exciton spectroscopy in MoTe$_2$ monolayers encapsulated in hBN.} T$=4$~K. (a) Differential reflectivity spectrum, the energy positions of the exciton transitions A:$1s$, A:$2s$ and B:$1s$ are marked. (b) Excitation with a HeNe Laser at 1.96~eV results in hot PL of the A:$2s$ and PL for the A:$1s$ state. The low energy peak labeled T might be related to the trion or phonon replica. (c) Photoluminescence excitation measurements detecting the emission from the A:$1s$ exciton. Peaks related to the resonant excitation of the A:$2s$ and B:$1s$ are marked. (d) Upconversion PL, the laser tuned into resonance with the A:$1s$ state results in emission about 120~meV higher energy, same as Fig.~\ref{fig:fig1}d.  }
\label{fig:fig3} 
\end{figure*}

\section{Excited state spectroscopy in ML $\mbox{MoTe}_2$}
\label{sec:mote2}
MoTe$_2$ is a very interesting layered material \cite{bie2017mote,jiang2017zeeman}, which provides the fascinating opportunity to switch between semiconducting $2H$ and metallic phases by tuning strain or carrier concentration \cite{Song:2015a,Li:2016a}. This allows working towards devices based on bias controlled phase changes in monolayer MoTe$_2$ \cite{Wang:2017mote2,doi:10.1021/acs.nanolett.6b04814}. First studies of $2H-$MoTe$_2$ flakes exfoliated on SiO$_2$ have identified monolayers as direct semiconductors \cite{Ruppert:2014a,Lezama:2015a}, interestingly the nature of the gap of the bilayer is still under discussion \cite{Robert:2016b,PhysRevB.94.085429}. So in practice the difference between mono- and bilayers has to be confirmed in Raman experiments, see Fig.~\ref{fig:fig4}a. MoTe$_2$ MLs have an optical bandgap at T$=4$~K at 1.17eV corresponding to an emission wavelength of 1050~nm. Therefore, its alloying with other TMD materials such as MoS$_2$ and MoSe$_2$ allows in principle to cover the full spectral range from 630 to 1050~nm for optoelectronics applications. As optical absorption is not only strong at the excitonic bandgap (A:1$s$) but also for higher lying exciton states \cite{Wang:2017b,He:2014a,Ugeda:2014a,Chernikov:2014a,Ye:2014a,Qiu:2013a,Ramasubramaniam:2012a,Wang:2015b} better knowledge of the excited exciton spectrum is needed. This allows also in principle to get an estimation of the exciton binding energy, by trying to compare with model calculations of exciton states in a screened 2D potential \cite{Rytova:1967,Keldysh:1979a,Chernikov:2014a,Robert:2018a}. \\
\indent Here we show the striking impact of hBN encapsulation on the optical properties of monolayer MoTe$_2$. The PL spectrum in Fig.~\ref{fig:fig3}b shows very narrow emission lines (FWHM linewidth of 3~meV) for the neutral exciton at the A:$1s$ state at 1.17~eV, approaching the optical quality reported for hBN encapsulated MoS$_2$ and WSe$_2$ monolayers \cite{Jin:2016a,Manca:2017a,Chow:2017a,Cadiz:2017a,Ajayi:2017a,Wang:2016v,wierzbowski2017direct}. We confirm the high sample quality in reflectivity experiments in Fig.~\ref{fig:fig3}a, that show this transition basically at the same energy as in PL, indicating negligible neutral exciton localization. In reflectivity we see also a broader transition about 250~meV above the A:$1s$ that we ascribe to the B:$1s$ state, following comparison with the data from the literature \cite{Ruppert:2014a,Lezama:2015a,yang2015robust}. We also observe in reflectivity a transition 120~meV above the A:$1s$ state, not reported previously, which we ascribe to the A:$2s$ state. Strikingly, when exciting with a laser energy of 1.96~eV we also see hot PL emission of this A:2$s$ transition in Fig.~\ref{fig:fig3}b.\\
\indent To further investigate the nature of these excited exciton states, we carry out PLE experiments. We monitor the PL emission of the A:$1s$ state (as in Fig.~\ref{fig:fig3}b) as a function of the laser excitation power. PLE probes absorption, which gives information on the higher lying electronic transitions, and subsequent relaxation to the A:$1s$ state, usually by emitting phonons. We observe in our experiments clear indications of both processes: absorption by excited exciton states and phonon assisted energy relaxation. In Fig.~\ref{fig:fig3}c, we see clear resonances in PLE exactly at the same energies as the reflectivity spectrum for the A:2$s$ and B:1$s$ state. The PL emission is enhanced by orders of magnitude where the laser is resonant with these excited exciton states, indicating efficient absorption and energy relaxation. More details on phonon assisted relaxation and associated Raman scattering on this sample are described in Appendix \ref{app:data}.\\
\indent As discussed in the previous section for MoSe$_2$ MLs, a powerful technique for investigating exciton states is photoluminescence upconversion. Here a \emph{cw} laser excites the MoTe$_2$ monolayer at the A:$1s$ resonance and emission at higher energies is monitored. In Fig.~\ref{fig:fig3}d we indeed observe emission 120~meV above the A:$1s$ state, this emission is exactly at the same energy as the transition ascribed to the A:$2s$ state with the three other spectroscopy techniques: reflectivity, hot PL and PLE all compared in Fig.~\ref{fig:fig3}a-d. \\
\indent To summarize the main experimental results, we have demonstrated that TMD MLs such as MoTe$_2$, MoSe$_2$, MoS$_2$ and WSe$_2$ exhibit strong photoluminescence upconversion. At resonant laser excitation of the A:$1s$ ground excitonic state luminescence from \emph{excited} states, such as A:$2s$ and B:$1s$ is detected. The effect vanishes for non-resonant excitation or if the oscillator strength of the A:$1s$ exciton is reduced by the gate-doping in charge tunable samples. Furthermore, the analysis of the upconversion PL intensity as a function of excitation power demonstrates that this effect is non-linear and requires two excitons in the ground state. In the next section we provide a theoretical model for our findings.

\section{Theory of exciton upconversion}

In this section we provide a theoretical model of the upconversion effect observed in TMD MLs. In this process the emitted photon energy is larger than that of the absorbed photon. That is why, in order to fulfil energy conservation, a third body, an exciton or a phonon should be involved. However, at a temperature of $4$~K the thermal phonon energies are less than $0.3$~meV, thus, lattice vibrations cannot provide effective transfer of excitons up to several  $100$~meV above the excitation energy. Additionally, experimental data shown in Fig.~\ref{fig:fig2}c (see also Ref.~\cite{Manca:2017a}) clearly demonstrates the presence of an optical nonlinearity: The upconversion intensity scales quadratically with the number of photoexcited excitons. Thus, in order to describe the upconversion theoretically we need to take into account exciton-exciton interaction processes where one of the excitons is annihilated while the second exciton acquires large extra energy, as depicted in the inset in Fig~\ref{fig:fig1}b \cite{Mouri:2014a,Kumar:2014b}. Subsequently, this exciton relaxes toward the radiative states (particularly, A:$2s$ and B:$1s$) and hot luminescence from these states is observed since the radiative recombination time is competitive (i.e. short enough) compared to the energy relaxation time.

This mechanism of generating  highly excited excitons can be viewed as Auger-like exciton-exciton annihilation. At first glance, it seems to be quite weak because in order to satisfy the energy and momentum conservation laws, the initial kinetic energy of the involved particles should be very large~\cite{abakumov_perel_yassievich}. As we show here this effect is very efficient in TMD MLs due to (i) strong Coulomb interaction, which makes it possible to relax the single-electron momentum conservation~\cite{PhysRevB.54.16625,PhysRevB.73.245424} and (ii) the possibility of a resonant processes involving exciton transfer to an excited energy band~\cite{Manca:2017a}. In other words, we try to include the particular energy spacing between different conduction bands in the theory of this four particle interaction.

\begin{figure}
\includegraphics[width=\linewidth]{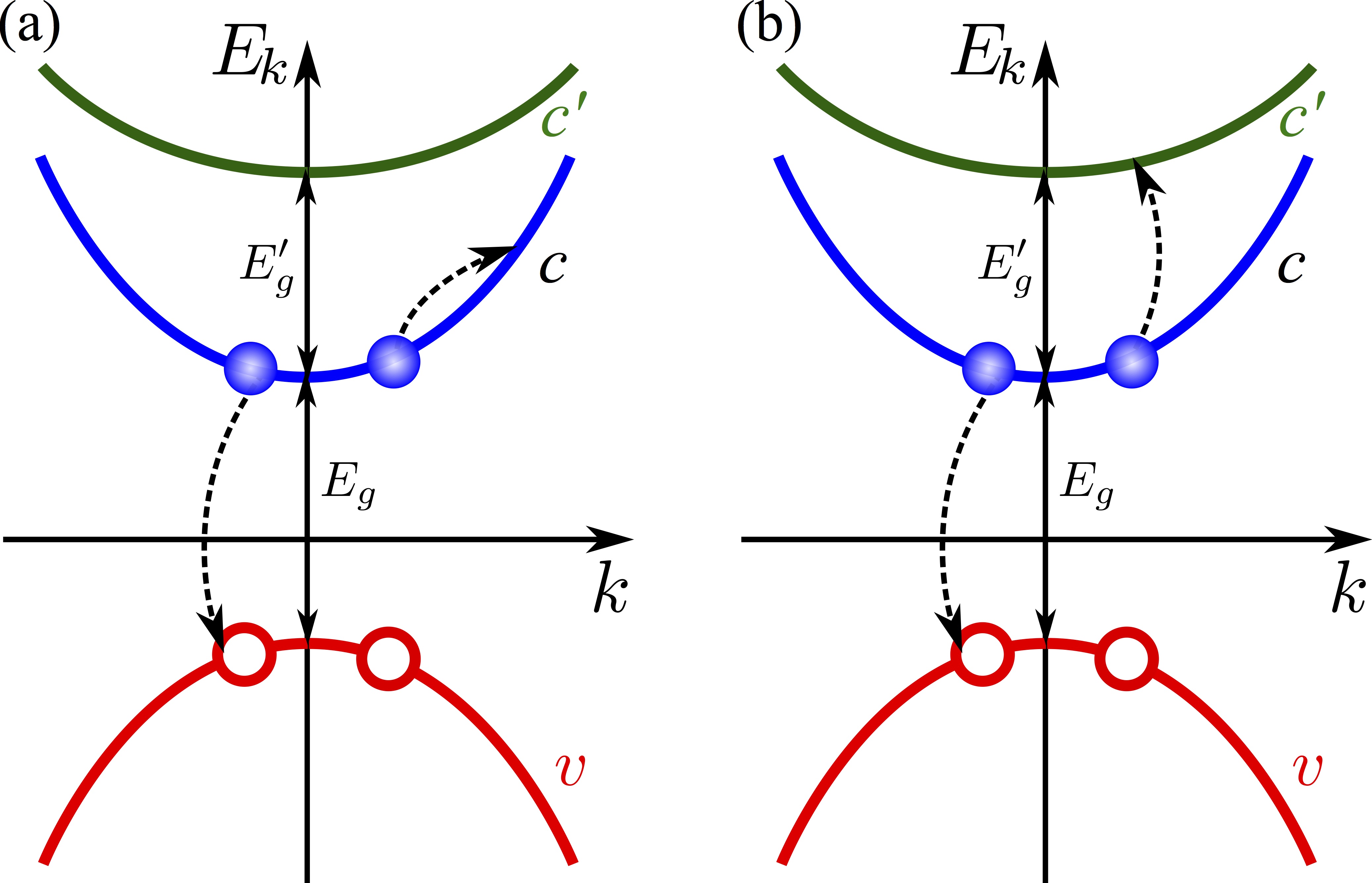}
\caption{\textbf{Model of exciton upconversion --- \textit{single particle picture}}. Filled circles denote electrons, open circled denote unoccupied states in the valence band. Dashed arrows show real electronic transitions.(a) Intraband process enabled by excitonic effects. The resulting high energy exciton involves carriers in the bands $c$ and $v$, the final momentum $\bm K_f=\bm K_1+\bm K_2$ is large. (b) Resonant Auger process resulting in high energy, low wavevector $\bm K_f$ excitons involving the excited conduction band $c'$. }
\label{fig:model} 
\end{figure}

The schematics of exciton-exciton Auger processes is presented in Fig.~\ref{fig:model}. Panel (a) shows an example of a standard Auger process which is possible in any semiconductor: Due to the Coulomb interaction one electron recombines with a hole, while another carrier is transferred to a highly excited state. Figure~\ref{fig:model}(b) illustrate a very different process, which is possible in the studied TMD MLs due to their specific band structure: It turns out that there is an excited conduction band (denoted as $c'$) whose distance to the conduction band, $E_g'$, approximately satisfies the condition
\begin{equation}
\label{cond}
E_g' \lesssim E_g - E_B,
\end{equation} 
where $E_B\approx 0.3\ldots 0.5$~eV~\cite{Wang:2017b} is the exciton binding energy, see Tab.~\ref{tab:gaps}. Thus, in the course of exciton-exciton annihilation the electron can be promoted to the $c'$ band with relatively small wavevector rather than be scattered to a large wavevector state within the same band. 

\begin{table}[b] 
\caption{Band gap energies. Data from DFT calculations summarized in Ref.~\cite{PhysRevB.95.155406} for S and Se-based MLs and from Ref.~\cite{Robert:2016b} for MoTe$_2$ MLs. $c'$ corresponds to $c+2$ band in notations of Refs.~\cite{PhysRevB.95.155406,Kormanyos:2015a}. The direct comparison of the values with experimental data in Fig.~\ref{fig:fig1} is not possible due to different levels of DFT approximations used. }
\label{tab:gaps}
\begin{tabular}{c|c|c|c|c}
\hline   
Energy (eV) & MoS$_2$ & MoSe$_2$ & WSe$_2$ & MoTe$_2$\\
\hline
$E_g$ & $1.8$ & $1.6$ & $1.7$ & $1.7$ \\
$E_g'$ & $1.1 \ldots 1.2$ & $1$ & $1.4$ & $1.3$\\
\hline
\end{tabular}
\end{table}

The rate of the Auger processes is described by a parameter $R_A$ such that the generation rate of highly energetic excitons is given by 
\begin{equation}
\label{Auger}
\frac{d n_{{X}}'}{dt} = R_A n_{{X}}^2.
\end{equation} 
Here $n_{{X}}'$ is the density of highly energetic excitons, $n_{{X}}$ is the density of photoexcited excitons in A:$1s$ state, recombination and energy relaxation processes are disregarded in Eq.~\eqref{Auger}. The same rate $R_A n_{{X}}^2$ describes the decay rate of A:$1s$ excitons due to the non-radiative exciton-exciton annihilation: $dn_{{X}}/dt = - R_A n_{{X}}^2$~\cite{abakumov_perel_yassievich,PhysRevB.54.16625,Kumar:2014b,Mouri:2014a,Sun:2014a,PhysRevB.93.201111,doi:10.1021/acs.jpclett.7b00885,PhysRevB.95.241403,PhysRevB.94.085429}. The rate $R_A$ can be expressed by means of the Fermi golden rule in the form
\begin{multline}
\label{RA}
R_A = \frac{2\pi}{n_{{X}}^2\hbar} \sum_{\bm K_1,\bm K_2, \nu} |M_{XX}|^2 f(K_1) f(K_2) \times \\{\delta[E_g -2E_B- E_g'+E(K_1) + E(K_2) - E_\nu(K_f)].}
\end{multline}
Here $M_{XX}$ is the matrix element of the exciton-exciton interaction, $\bm K_1$ and $\bm K_2$ are the center of mass wavevectors of the two interacting excitons, $f(K)$ is the distribution function of photoexcited A:$1s$ excitons, $E(K)=\hbar^2 K^2/2M$ is the exciton dispersion with $M$ being its effective mass, the subscript $\nu$ denotes the quantum numbers of the final exciton state which include the electron band index [$c$ or $c'$ for the processes shown, respectively, in Fig.~\ref{fig:model}(a) or (b)] as well as of the internal motion ($2s$, $2p$, \ldots including the continuum states), $E_\nu(K)$ is the dispersion of the exciton in the final state, which accounts for its binding energy. The exciton wavevector in the final state, $\bm K_f$ is found from the momentum conservation law: $\bm K_f = \bm K_1 + \bm K_2$. For the derivation of Eq.~\eqref{RA} we assumed that the occupation of the final states is negligible and omitted the corresponding occupation factor and also disregarded the anisotropy and nonparabolicity of exciton dispersion.

\begin{figure}[t]
\includegraphics[width=\linewidth]{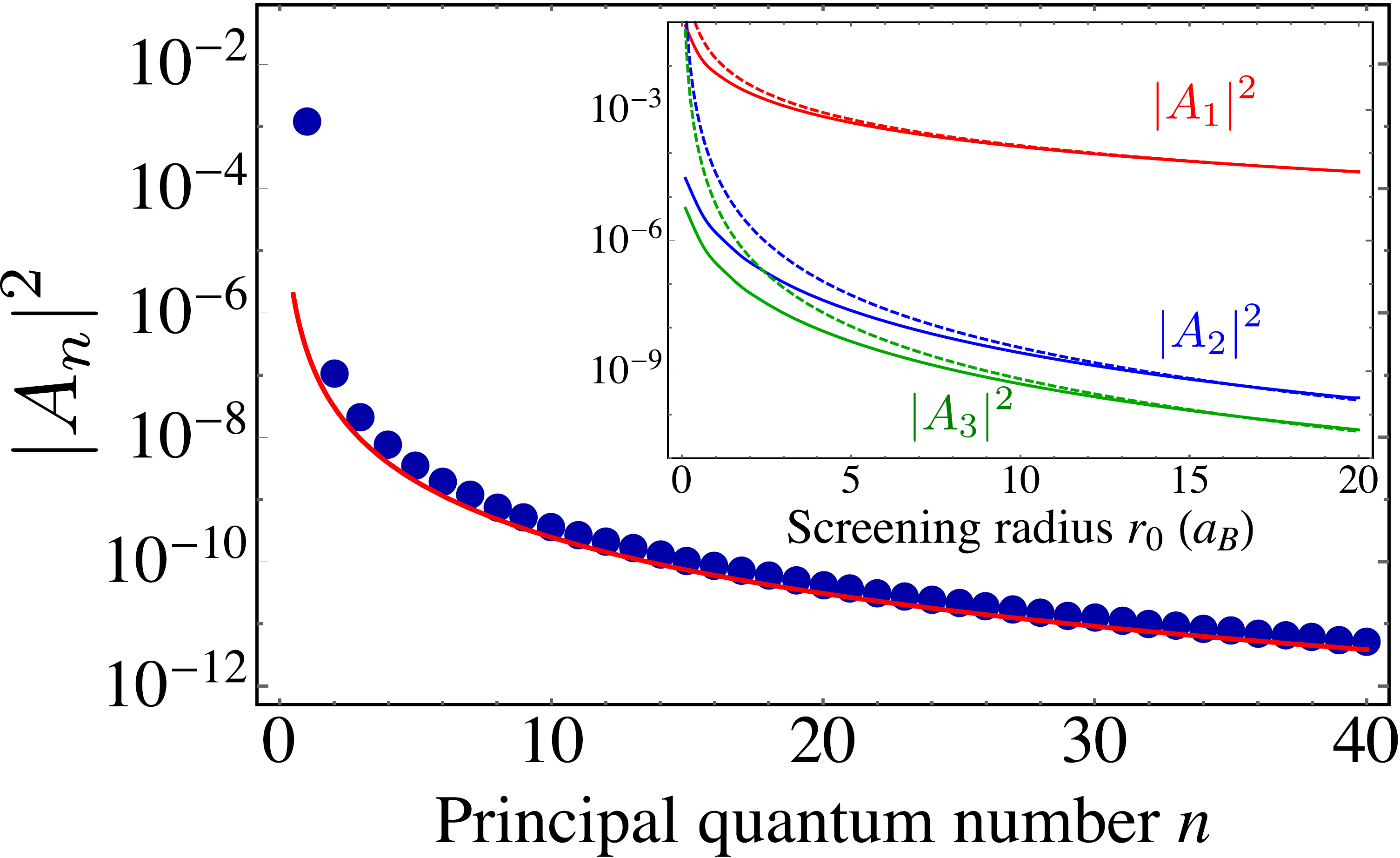}
\caption{Dependence of the coefficient $|A_n|^2$ on $n$ for $r_0=3a_B$. Red line shows the approximation $|A_n|^2 =2.47\times 10^{-7}/n^3$. Inset shows dependences $|A_{1}|^2$ (red), $|A_{2}|^2$ (blue) and $|A_{3}|^2$ (green) on the screening radius $r_0$. Dashed lines are fits $|A_{1}|^2 = 0.015(a_B/r_0)^{2}$ and $|A_{2}|^2 = 3.46\times 10^{-5}(a_B/r_0)^{4}$, $|A_{3}|^2 = 6.63\times 10^{-6}(a_B/r_0)^{4}$.}
\label{fig:A_n}
\end{figure}

The analytical results are obtained in Appendix~\ref{app:theory} under the model assumptions $E_B \ll E_g, E_g'$, which allow us to make use of $\bm k \cdot \bm p$-perturbation theory for calculating the excitonic states and transition rates~\cite{Glazov:2017a}. The analysis demonstrates that the dominating contribution to $R_A$ is given by the resonant processes described in Fig.~\ref{fig:model}b where the electron in the exciton is promoted to the excited band $c'$. The Coulomb interaction between the electrons can result in the process shown in Fig.~\ref{fig:model}(b) where one pair recombines while the remaining electron occurs in the $c'$ band. So, one $c'v$ pair is present in the end. Since in this process both electrons change their quantum states the resonant Auger scattering is due to the electron-electron interaction only, while the electron-hole interaction does not play a role. Moreover, our analysis shows that the exchange contribution to the matrix element dominates, see Appendix~\ref{app:theory}. 

Our next target is to evaluate the scattering rate for exctions under the resonant condition $E_g=E_g'+2E_B-E_{B,n}$, where the energy released at the non-radiative recombination of A:$1s$ exciton with $\bm K=0$ is equal to the energy of the excited $c'$-band exciton in the $ns$-state. Neglecting the difference of exciton masses in the initial and final states and assuming that excitons are thermalized with the temperature $T$ we have 
\begin{equation}
\label{RA:res}
R_A = \frac{\pi}{\hbar k_B T} \left|\frac{2\pi e^2 \gamma_3 \gamma_6 }{{\varkappa} a_B^2 E_gE_g'}\right|^2 |A_n|^2.
\end{equation} 
Here $k_B$ is the Boltzmann constant, $a_B$ is the exciton Bohr radius $\gamma_3$ and $\gamma_6$ are the interband momentum matrix elements (in the units of $m_0/\hbar$, $m_0$ being the free electron mass) for electron transition from $c$ to, respectively, $v$ and $c'$ bands, $\varkappa$ is the effective high-frequency dielectric constant. In Eq.~\eqref{RA:res} $A_n$ is the dimensionless overlap integral which depends on the screening parameter $r_0$ of the Coulomb potential given by the Fourier transform $V_C(q) = {2\pi e^2/[\varkappa q(1+qr_0)]}$, see Appendix~\ref{app:theory} for details. The values of $|A_n|^2$ for several excitonic states are shown in Fig.~\ref{fig:A_n}. For $n=1$ and reasonable material parameters~\cite{PhysRevB.95.155406} the quantity $R_A$ at $T=4$~K in Eq.~\eqref{RA:res} can be estimated to be $1\ldots 10$~cm$^2$/s. This quantity can be reduced by a factor $10\ldots 100$ if the resonant condition is not fulfilled. The rate of intraband transitions, where the electron remains in the same band, Fig.~\ref{fig:model}(a), can be estimated by replacing the factor $\gamma_6/(a_B E_g')$ by $(E_B/E_g)^{3/2}$ which produces a parametrically smaller, $\sim E_B/E_g$ contribution, see Appendix~\ref{app:theory}. In the non-resonant case where at $ K_1= K_2=0$ the resonant transition is not possible and the detuning $\Delta = E_g'-E_g + 2E_B - E_{B,n}$ is present for the intraband process. This intraband process is a only possible when excitons with the kinetic energy on the order of $\Delta$ are present. The rate Eq.~\eqref{RA:res} acquires therefore an exponential factor $\exp{(-|\Delta|/k_B T)} < 1$. 

The experimentally observed Auger rates in the literature are one-two orders of magnitude smaller than the resonant contribution to $R_A$ investigated here~\cite{Kumar:2014b,Mouri:2014a,Sun:2014a,PhysRevB.93.201111,doi:10.1021/acs.jpclett.7b00885,PhysRevB.95.241403,Robert:2016a}.  The exact values of $R_A$ will also vary with sample temperature and environment, demonstrating that the exact resonance conditions are not fulfilled in the studied structures. Our experiments are carried out at 4~K whereas many exciton-exciton scattering studies are carried out at elevated temperatures. The presence of disorder in the sample, especially without hBN encapsulation, may enable to fulfil simultaneously the energy and momentum conservation in TMD MLs making additional scenarios possible.\\
\indent To summarize, our analysis suggests that exciton upconversion photoluminescence in TMD MLs is  due to a specific nonlinear  process: Two excitons generated by the resonant laser collide, as a result one of those recombines non-radiatively while the other is promoted to a highly excited state (most likely related to an excited conduction subband according to band structure calculation). Subsequently, the excited exciton loses its energy and a hot PL from the radiative A:$2s$ and B:$1s$ states is observed. This scenario describes the main experimental findings: (i) upconversion PL from the states A:$2s$ and B:$1s$ which are visible in the hot PL [Figs.~\ref{fig:fig1},~\ref{fig:fig2}(a,b,d) and~\ref{fig:fig3}], (ii)~quadratic dependence of the upconversion intensity on the number of excitons in the ground state, Fig.~\ref{fig:fig2}(c), (iii) resonant character of the process as a function of excitation laser energy, Fig.~\ref{fig:fig2}(d), and (iv) absence of the upconversion in the presence of doping, where the exciton resonance vanishes, Fig.~\ref{fig:fig2}(f).

\section{Conclusion}
We identify  of excited exciton states in high quality MoSe$_2$ and MoTe$_2$ monolayer samples, which govern absorption and emission above the A:1$s$ exciton resonance. We identify the A:2$s$ state 150~meV (120~meV) above the A:1$s$ state in ML MoSe$_2$ (MoTe$_2$). We show that excited exciton states can be studied in photoluminescence upconversion experiments. In addition to being a highly selective spectroscopic tool applicable to several TMD materials \cite{Manca:2017a,Robert:2018a}, this non-linear optical effect also gives insights into exciton-exciton interactions, relevant physical processes also for studying population inversion and other density dependent phenomena \cite{Li:2017a,Wu:2015b,Chernikov:2015z,PhysRevB.95.241403}. Our work suggests that in TMD monolayers the generation of high energy excitons with Auger-like scattering processes is efficient due to the strong Coulomb interaction and resonant excitation of higher lying conduction bands.

\begin{acknowledgements}  
We acknowledge funding from ANR 2D-vdW-Spin, ANR VallEx, Labex NEXT projects VWspin and MILO, ITN Spin-NANO Marie Sklodowska-Curie grant agreement No 676108 and ITN 4PHOTON Nr. 721394. X.M. also acknowledges the Institut Universitaire de France. Growth of hexagonal boron nitride crystals was supported by the
Elemental Strategy Initiative conducted by the MEXT, Japan and the CREST
(JPMJCR15F3), JST. 
L.E.G. and M.M.G. acknowledge partial support from LIA ILNACS, RFBR projects 17-02-00383, 17-52-16020, RF President Grant MD-1555.2017.2 and ``BASIS'' foundation.
\end{acknowledgements}


\appendix

        \setcounter{figure}{0}
        \renewcommand{\thefigure}{S\arabic{figure}}%

\section{Experimental Methods}\label{app:A}
The samples are fabricated by mechanical exfoiliation of bulk MoSe$_2$ and MoTe$_2$ (commercially available from 2D semiconductors) and very high quality Hexagonal Boron Nitride (hBN) crystals \cite{Taniguchi:2007a} on 83 nm SiO$_2$ on a Si substrate. The experiments are carried out at T = 4~K in a confocal microscope built in a vibration free, closed cycle cryostat. The excitation/detection spot diameter is $\sim 1 \mu m$. 
The monolayer (ML) is excited by continuous wave Ti-Sa laser (700-1020~nm) or a HeNe laser (633~nm). The photoluminescence (PL) signal is dispersed in a spectrometer and detected with a Si-CCD camera ($\lambda < 1\mu m$) or InGaAs detector ($\lambda > 1\mu m$). The typical excitation power is 3 $\mu W$.

\section{Additional data}\label{app:data}

\begin{figure}[h]
\includegraphics[width=0.45\textwidth,keepaspectratio=true]{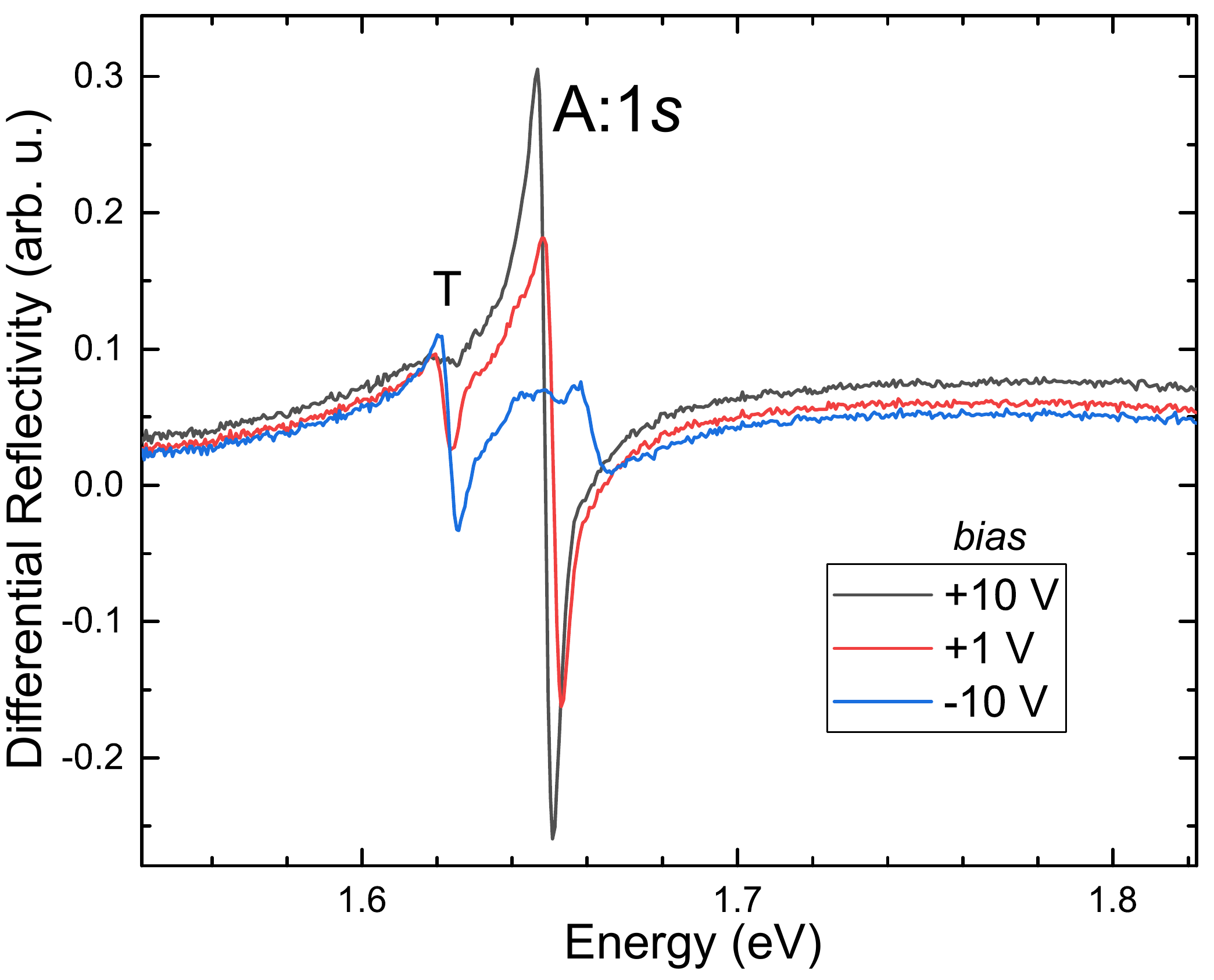}
\caption{\textbf{Charge tunable device} In differential reflectivity on the charge tunable device of Fig.~\ref{fig:fig2}(e,f) we identify the A:1$s$ state and at more negative bias the charged exciton state marked T for trion. }
\label{fig:reflbias} 
\end{figure}

\textbf{Charge tunening in ML MoSe$_2$. }Figure~\ref{fig:reflbias} demonstrates reflectivity spectrum of the charge tunable MoSe$_2$ device clearly showing the redistribution of the oscillator strength between the neutral and charged excitons, studied in the context of upconversion in Fig.~\ref{fig:fig2}(e,f).

\begin{figure*}
\includegraphics[width=0.9\textwidth,keepaspectratio=true]{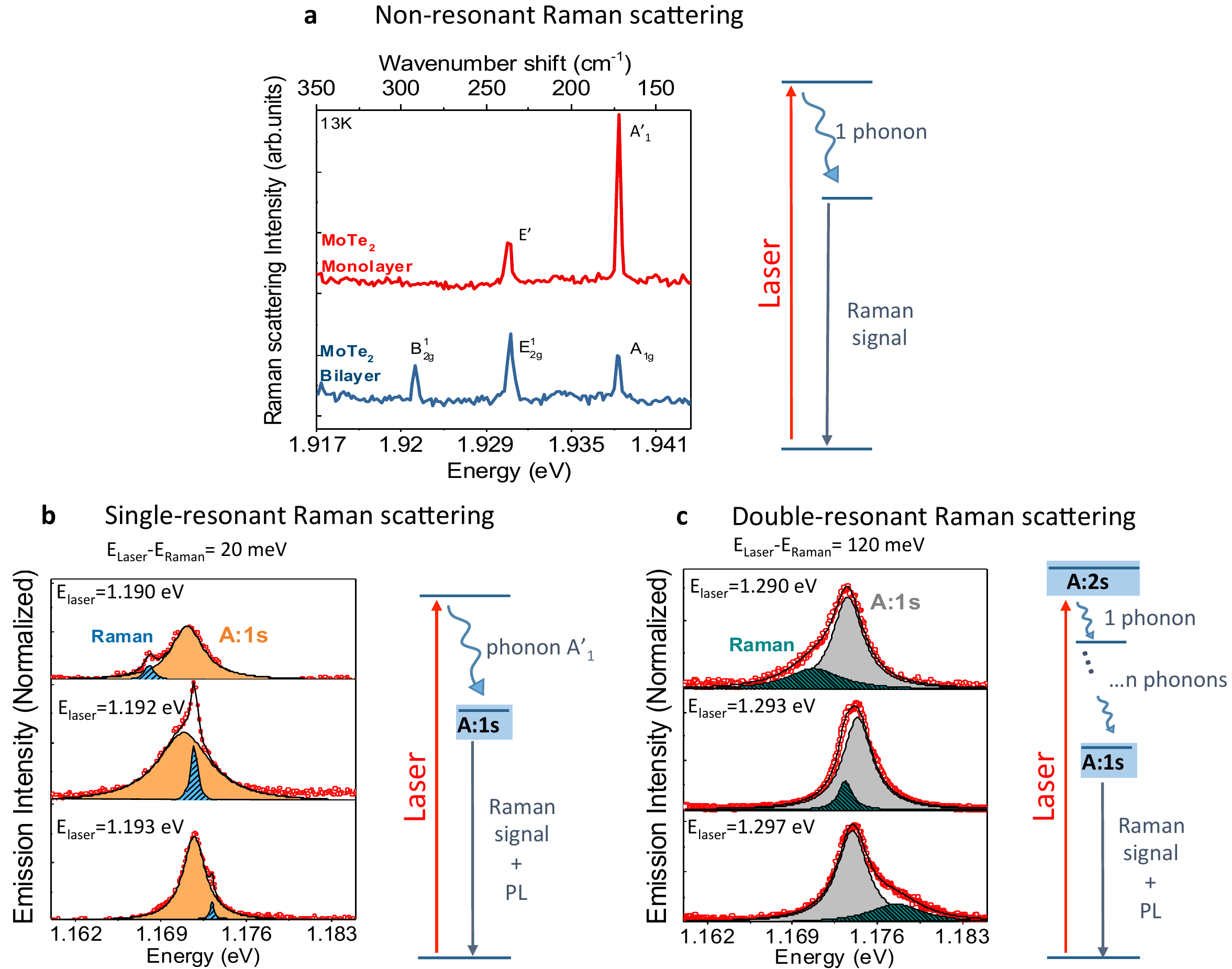}
\caption{\textbf{Raman spectroscopy in ML MoTe$_2$. } $T=4$~K. (a) Non-resonant Raman scattering using a HeNe laser. (b)  Single resonant Raman scattering as the excitation laser energy is one phonon energy $A'_{1}$ above the A:$1s$ state. (c) Double resonant Raman experiments as a phonon multiple ensures efficient relaxation from the optically excited A:$2s$ state to the emitting A:$1s$ state. }
\label{fig:fig4} 
\end{figure*}
\textbf{Resonant and non-resonant Raman scattering in ML MoTe$_2$.} Scattering with phonons in Raman processes allows to distinguish monolayers from multilayers. This is especially useful for MoTe$_2$ where also the bilayer shows clear and narrow PL emission. In Fig.~\ref{fig:fig4}a we compare results for a monolayer and a bilayer, the absence of the $B^1_{2g}$ peak allows us to identify monolayer samples \cite{Guo:2015a,tonndorf:2013a}. The experiments in Fig.~\ref{fig:fig4}a are based essentially on non-resonant Raman scattering i.e. neither the laser energy nor the emitted light after phonon scattering are resonant with an particular electronic state. This is different in Fig.~\ref{fig:fig4}b : Here we tune the laser to an excess energy of about 20~meV above the A:1$s$ resonance. In addition to PL emission (orange peak) we see a spectrally sharper feature (shaded blue) superimposed on the PL, which shifts with excitation laser energy. This peak is corresponds to Raman scattering with the $A'_{1}$ phonon, which is particularly efficient as the final state after scattering corresponds to a real electronic state ~\cite{Guo:2015a,Carvalho:2015a,Wang:2015g,Chow:2017a,molas2017raman,Soubelet:2016a} in these single-resonant Raman scattering experiments.\\
\indent In Fig.~\ref{fig:fig4}c we report double resonant Raman experiments~\cite{Wang:2015g}. As the laser energy is scanned across the A:$2s$ state, we see that the PL of the A:$1s$ exciton is enhanced, see intensity plotted in Fig.~\ref{fig:fig3}c as a function of laser energy. In addition, we 
observe in Fig.~\ref{fig:fig4}c that a Raman feature is crossing the PL line, exactly 120~meV below the respective laser energy. When the laser is at the A:$2s$ energy, the Raman process is double resonant \cite{Guo:2015a,Wang:2015g} as the initial state (A:$2s$) and the final state (A:$1s$) are real electronic states. This has already been observed between exciton states in ML WSe$_2$ on SiO$_2$ \cite{Wang:2015g}. Please note that very different electronic states and phonons are discussed in the double resonant Raman experiments in Ref.~ \cite{Guo:2015a}. In experiments in hBN encapsulated ML WSe$_2$ samples the similar experiments have been interpreted as being due to phonon related processes only \cite{Jin:2016a}. This interpretation seems unlikely in view of follow-up studies of ML WSe$_2$ in magnetic fields, that clearly showed that the excited state is an excitonic transition and not just phonon replica \cite{Stier:2018a}. For the case of MoTe$_2$ we have a strong case for the transition at 120~meV above the A:$1s$ to be attributed to a real electronic transition, as this transition is confirmed in Fig.~\ref{fig:fig3}a-d by four complementary spectroscopy techniques and in Fig.~\ref{fig:fig4}c by double resonant Raman scattering. \\

\section{Calculations of the Auger rates}\label{app:theory}

\subsection{Resonant interband process}

We consider three bands, $c,v,c'$, schematically illustrated in Fig.~\ref{fig:model} with two excited electron-hole pairs (excitons) with an electron occupying the lowest conduction band $c$ and with an empty state in the valence band $v$. Here for simplicity we disregard the spin degree of freedom of change carriers, assuming that the spin is conserved in the course of exciton-exciton interaction. Furthermore, we focus on the states in the vicinity of one of the band extrema ($\bm K_+$ or $\bm K_-$ valley) and disregard here the processes involving intervalley transfer of electron-hole pair studied in Refs.~\cite{Glazov:2014a}. The Coulomb interaction between the electrons can result in the process shown in Fig.~\ref{fig:model}(b) where one pair recombines while the remaining electron occurs in the $c'$ band. So, one $c'v$ pair is present in the end. Since in this process both electrons change their quantum states the resonant Auger scattering is due to the electron-electron interaction only, while, e.g., electron-hole interaction does not play a role.

In the free-particle picture, we have two electrons in the $c$ band which occur in the $c'$- and $v$-bands after the Coulomb scattering. The two-electron wavefunctions of the considered system in the initial and final states can be presented as
\begin{align}
\label{psi:ee}
&	\left| i \right > =  \frac{1}{\sqrt{2}}[\Psi_{c \bm k_c}(\bm r_1)\Psi_{c \widetilde{\bm k}_c}(\bm r_2) - \Psi_{c \bm k_c}(\bm r_2)\Psi_{c \widetilde{\bm k}_c}(\bm r_1)],
	\\
&	\left| f \right > = \frac{1}{\sqrt{2}}[\Psi_{v \bm k_v}(\bm r_1)\Psi_{c' {\bm k}_{c'} }(\bm r_2)- \Psi_{v \bm k_v}(\bm r_2)\Psi_{c' {\bm k}_{c'} }(\bm r_1)].\nonumber
\end{align}
Here $\bm k_c$, $\bm k_v$ are the electron and unoccupied state wavevectors in one of the excitons and $\widetilde{\bm k}_c$, $\widetilde{\bm k}_v$ are the electron and unoccupied state wavevectors in another exciton. The wavefunction in each band $n=c,c',v$ is a product of the Bloch amplitude and the plane wave:
\begin{equation}
	\Psi_{n \bm k}(\bm r) = \text{e}^{i \bm k \cdot \bm r} u_{n \bm k}(\bm r),
\end{equation}
and the normalization area is set to unity.
In the $\bm k \cdot \bm p$ model, the Bloch amplitudes have the form:
\begin{align}
	u_{c \bm k_c} = u_c + {\hbar \over m_0} {\bm k_c \cdot \bm p_{vc} \over E_c - E_v} u_v + {\hbar \over m_0} {\bm k_c \cdot \bm p_{c'c} \over E_c - E_{c'}} u_{c'}, \\
	u_{c' \bm k_{c'}} = u_{c'} + {\hbar \over m_0} {\bm k_{c'} \cdot \bm p_{vc'} \over E_{c'} - E_v} u_v  + {\hbar \over m_0} {\bm k_{c'}\cdot \bm p_{cc'} \over E_{c'} - E_c} u_c, \\
	u_{v \bm k_v} = u_v + {\hbar \over m_0} {\bm k_v \cdot \bm p_{cv} \over E_v - E_c} u_c + {\hbar \over m_0} {\bm k_v \cdot \bm p_{c'v} \over E_v - E_{c'}} u_{c'},
\end{align}
where $u_n$ denotes the Bloch amplitude at the extremum point, $m_0$ is the free electron mass and $\bm p_{nn'}$ are the momentum matrix elements between the states in the bands $n$ and $n'$ ($n,n'=c,c',v$).

The wavefunctions $|i\rangle$, $|f\rangle$ in Eq.~\eqref{psi:ee} are  antisymmetrized with respect to the permutations of electrons. It gives rise to the \emph{direct} and \emph{exchange} contributions. The matrix element of the direct interaction, where the electron from the state with the wavevector $\bm k_c$ recombines with the hole from same exciton and transfers to the state $\bm k_v$, can be conveniently presented in the form
\begin{multline}
	M_{dir} = \left<u_{v \bm k_v}| u_{c \bm k_c}\right>\left<u_{c' \bm k_{c'}}| u_{c \widetilde{\bm k}_c}\right> \times \\
	V_C(K_1)
	\delta_{\bm K_1, \bm k_c - \bm k_v}\delta_{\bm K_1, \bm k_{c'}-\widetilde{\bm k}_c},
\end{multline}
where $\bm K_1 = \bm k_c - \bm k_v$ is the exciton center of mass momentum (note that the hole state corresponds to the time-reversed counterpart of the unoccupied state)
\begin{equation}
\label{Coulomb}
V_C(q) = {2\pi e^2\over \varkappa q(1+qr_0)}
\end{equation} 
is the 2D Fourier image of the Coulomb potential with $\varkappa$ being the background average constant of the surrounding structure and $r_0$ being the dielectric screening parameter~\cite{Rytova:1967,Keldysh:1979a,Cudazzo:2011a}. Note that this parameter should be taken in the high-frequency limit because the energy transferred in the course of exciton-exciton interaction is on the order of the band gap $E_g$.

Taking into account that
\begin{equation}
\label{kp:elem}
{\hbar \over m_0} \bm k \cdot \bm p_{vc}=\gamma_3^* k_+, \quad {\hbar \over m_0} \bm k \cdot \bm p_{c'c}=\gamma_6 k_-,
\end{equation}
where $\gamma_3$ and $\gamma_6$ are the band structure parameters introduced in Refs.~\cite{Kormanyos:2015a,PhysRevB.95.155406} and $k_\pm = {k_x\pm \mathrm i k_y}$,
 we obtain:
\begin{equation}
	\left<u_{v \bm k_v}| u_{c \bm k_c}\right> = {\gamma_3^* K_+ \over E_c - E_v},
	\quad
	\left<u_{c' \bm k_{c'}}| u_{c \widetilde{\bm k}_c}\right>= 
	{\gamma_6 K_- \over E_{c'} - E_c}.
\end{equation}
Finally, the direct interaction matrix element takes a simple form
\begin{equation}
\label{dir:free}
	 M_1(K_1) \equiv V_C(K_1)  
	{\gamma_3^* \gamma_6 K_1^2 \over E_gE_g'} \delta_{\bm K_1, \bm k_c - \bm k_v}\delta_{\bm K_1, \bm k_{c'}-\widetilde{\bm k}_c}.
\end{equation}
With account for the excitonic effect Eq.~\eqref{dir:free} should be averaged over the exciton wavefunction~\cite{birpikus_eng,Glazov:2014a,Glazov:2017a}. Furthermore, we need to take into account that in the initial state there are two unoccupied states in the valence band.
As a result, we have ($K_1,K_2 \ll a_B^{-1}$)
\begin{multline}
\label{dir:exc}
	M_{dir}(\bm K_1, \bm K_2, \bm K_f,n) = \delta_{\nu,1s} \delta_{\bm K_f, \bm K_1+\bm K_2} \Phi_{1s}(0) \times \\
	\frac{1}{2}\left[V_C(K_1)K_1^2 + V_C(K_2)K_2^2\right] { 	\gamma_3^* \gamma_6  \over E_g E_g'}
	. 
\end{multline}
We recall that $\bm K_1$, $\bm K_2$ are the wavevectors of excitons in the initial state, $\bm K_f = \bm K_1+\bm K_2$ is the wavevector of the exciton in the final state, the subscript $\nu$ enumerates the relative motion states of the remaining electron-hole pair. In derivation of Eq.~\eqref{dir:exc} we neglected the difference of electron effective masses in $c$ and $c'$ bands and assumed that initially both excitons occupy $1s$ state, $\Phi_{1s}(\rho)$ is the envelope function of the relative motion. Correspondingly, the final state relative motion envelope function remains the same.

Typical center of mass wavevectors involved in exciton-exciton scattering are on the order of thermal wavevector $K_T =\sqrt{2M k_B T/\hbar^2}$ and are much smaller than the screening wavevector $r_0^{-1}$, therefore the direct exciton-exciton scattering matrix element $M_{dir}$ is proportional to the first powers of the exciton wavevectors: $M_{dir} \propto K_T$.


Here we consider an exchange process, where the electron occupies the empty state in the valence band related to the hole in the other exciton, i.e. the electron with the wavevector $\bm k_c$ transfers to the valence band state with the wavevector $\tilde{\bm k}_v$. As a result for uncorrelated electron-hole pairs we have for the exchange contribution
\begin{equation}
\label{exc:free}
 -M_1(|\bm k_c - \widetilde{\bm k}_v|),
\end{equation}
where $M_1$ is defined in Eq.~\eqref{dir:free}.	In order to transform Eq.~\eqref{exc:free} to the form convenient for averaging over the exciton wavefunctions we  introduce the relative motion wavevectors for two initial and final exciton states in accordance with 
\begin{equation}
 \bm k_1 = {\bm k_c+\bm k_v\over 2},
	\qquad	
	 \bm k_2 = {\widetilde{\bm k}_c+\widetilde{\bm k}_v\over 2},
\end{equation}
\[
 \bm k_f = {\bm k_{c'}+\bm k_v\over 2} = \bm k_1 + {\bm K_2\over 2}.
\]
Here we assumed that the effective masses of the electron and hole are the same in agreement with microscopic calculations~\cite{Kormanyos:2015a,PhysRevB.95.155406}.
Taking into account that, as before, the center of mass wavevectors $K_1,K_2 \sim K_T$ are small compared with the inverse Bohr radius $a_B^{-1}$ of exciton we omit $\bm K_f$ in $\bm k_c - \widetilde{\bm k}_v = \bm k_1 - \bm k_2 +\bm K_f/2$ and in $\bm k_1+\bm K_f/2$ and
arrive at 
\begin{multline}
\label{M_exch}
	M_{exch}(\bm K_1, \bm K_2, \bm K_f,\nu)  	\approx  - \delta_{\bm K_f, \bm K_1+\bm K_2} \times \\
	\sum_{\bm k_1, \bm k_2} M_1\left(|\bm k_1-\bm k_2 |\right) C_{\nu}^*\left(\bm k_1 \right) C_{1s}( k_2)  C_{1s}( k_1) . 
\end{multline}
Here $C_\nu (\bm k)$ are the Fourier transforms of the relative motion exciton functions $\Phi_{\nu}(\bm \rho)$:
\[
C_\nu (\bm k) = \int d\bm \rho \, \mathrm e^{\mathrm i \bm k \cdot \bm \rho} \,\Phi_\nu(\bm \rho).
\]
It follows from Eq.~\eqref{M_exch}  that only $s$-shell states contribute to the matrix element. As compared with its direct counterpart, the transferred momentum here is $|\bm k_1-\bm k_2 |\sim a_B^{-1}$.  Since $M_1(q) \propto q$ for $qr_0 \ll 0$, the direct contribution is by a factor $K_T a_B$ smaller than the exchange one. Thus, in what follows we consider the exchange contribution only.

In order to analyze the exchange process in more detail we first consider a limit where the screening is very strong, i.e., where $|\bm k_1-\bm k_2| r_0 \gg 1$. In this case we can approximate $M_1(q)$ by a constant and arrive at
\begin{multline}
\label{M_exch_inter}
	M_{exch}(\bm K_1, \bm K_2, \bm K_f,\nu)  \approx  \\
	- \delta_{\nu,1s} \delta_{\bm K_f, \bm K_1+\bm K_2}  {2\pi e^2\over \varkappa r_0} {\gamma_3^* \gamma_6 \over E_gE_g'} \Phi_{1s}(0).
\end{multline}

For arbitrary screening we evaluate the sum in Eq.~\eqref{M_exch} making use of the two-dimensional hydrogenic functions. For the bound states $\nu = ns$ we have~\cite{chao:6530}
\begin{equation}
\label{C_ns}
C_{ns}(k) =  2\sqrt{2\pi}a_B  \left( {2n-1 \over 1+\kappa_n^2} \right)^{3/2} P_{n-1} \left({\kappa_n^2-1 \over \kappa_n^2+1} \right),
\end{equation}
with $\kappa_n=(2n-1)ka_B$ and $P_n(x)$ being the Legendre polynomial. As a result, 
\begin{equation}
	M_{exch}(\bm K_1, \bm K_2, \bm K_f,ns)  = \frac{A_n(r_0)}{a_B^2} {2\pi e^2\gamma_3^* \gamma_6 \over \varkappa E_gE_g'}  \delta_{\bm K_f, \bm K_1+\bm K_2}.
\end{equation}
Dependence of $|A_n|^2$ on $n$ for $r_0=3a_B$ is shown in Fig.~\ref{fig:A_n}. We see that the squared matrix element decreases rapidly with $n$. It follows from Eq.~\eqref{C_ns} that $C_{n0} \propto n^{-3/2}$ at $n \to \infty$, therefore $|M_b|^2 \sim n^{-3}$. Figure~\ref{fig:A_n} shows that this asymptotic is valid already at $n \geq 3$.

The inset to Fig.~\ref{fig:A_n} shows the dependences of the scattering probability on the screening radius $r_0$. Final states with $n=1,2,3$ are considered. As it is mentioned above, scattering into the $ns$ state for the short-range interaction is possible at $n=1$ only. The probability of this process decreases as $1/r_0^2$, while for $n\geq 2$ it drops as $1/r_0^4$. The corresponding asymptotes are shown by dashed lines in Fig.~\ref{fig:A_n}.

\subsection{Intraband Auger process}

Let us now briefly address the intraband process depicted in Fig.~\ref{fig:model}(a) where the charge carriers remain in the same bands after the scattering. For free carriers the process has a high threshold requiring the initial and final wavevectors to be on the order of $\sqrt{M E_g/\hbar^2}$, otherwise the energy and momentum conservation laws cannot be satisfied simultaneously. With account for the excitonic effect the process becomes allowed because one can find, in the relative motion wavefunction Fourier image, Eq.~\eqref{C_ns}, sufficiently large wavevectors due to the Coulomb interaction. In other words, the electron-hole Coulomb interaction either in the initial or in the final state may relax the momentum conservation in the course of the Auger scattering. However, the two band approximation is insufficient to give a correct result, since (i) the model approximations for the band dispersions are, as a rule, invalid at the kinetic energies $\sim E_g$ due to the $\bm k\cdot \bm p$ interaction with remote bands~\cite{abakumov_perel_yassievich,Kormanyos:2015a} and (ii) the asymptotic form of $C_{\nu}(k)$ at large wavevectors can strongly differ from a simplified hydrogenic model~\eqref{C_ns}~\cite{birpikus_eng,Qiu:2013a,Wang:2017b}. Thus, we present here only analytical estimations based on the parabolic approximations for the band dispersions and assuming that the ratio $E_g/E_B$ is very large, which allows to take into account the Coulomb effects perturbatively.

We start from the direct process. Instead of Eq.~\eqref{dir:free} we have for the free carrier scattering 
\begin{equation}
 V_C(K_1){\gamma_3^* K_{1,+} \over E_g }\delta_{\bm K_1, \bm k_c - \bm k_v}\delta_{\bm K_1, \bm k_{c'}-\widetilde{\bm k}_c}.
\end{equation}
As compared with Eq.~\eqref{dir:free} the factor $\propto \gamma_6/E_g'$ is absent due to the fact that only one charge carrier changes the band. Making use of the following notations
\begin{align}
\bm k_{c'} = \bm k_f + {\bm K_1 + \bm K_2\over 2} = \bm k_f + {\bm K_f\over 2},\\
\widetilde{\bm k}_c = \bm k_f + {\bm K_2 - \bm K_1\over 2},\nonumber
\end{align}
we obtain for the exciton Auger scattering matrix element the following expression
\begin{multline}
\label{M_direct_intra}
	M_{dir}'(\bm K_1, \bm K_2, \bm K_f, \bm k_f) = 	\delta_{\bm K_f, \bm K_1+\bm K_2} \\
	\times \Phi_{1s}(0){V_C(K_1) 	\gamma_3^* K_{1,+} \over E_g}
	\int d\bm r \text{e}^{\mathrm i \bm K_1 \cdot \bm r/2} \Phi_{k_f,l}^*(\bm r) \Phi_{1s}(\bm r) \\
	+ \bm K_1 \leftrightarrow \bm K_2. 
\end{multline}
Here we take into account that only excitons with positive energies of relative motion can be in the final state, i.e., $\bm k_f$ corresponds to the continuum electron-hole pair state modified by the Coulomb interaction. The final state wavevector can be estimated from the energy conservation condition with the result $k_f = \sqrt{(E_g - 2E_B)M/(2\hbar^2)}$. In this estimate we neglected thermal energy of excitons as compared with $E_g - 2E_B$ and, as before, took the same effective masses for an electron and a hole.

Since $K_1 \ll a_B^{-1}, k_f$ the integral in Eq.~\eqref{M_direct_intra} can be evaluated decomposing the exponent in the series. At $K_1=0$ the integral in Eq.~\eqref{M_direct_intra} equals to zero due to orthogonality of the functions of discrete and continuous spectra. Therefore we take into account the $K_1$-linear term:
\begin{equation}
\int d\bm r \text{e}^{\mathrm i \bm K_1 \cdot \bm r/2} \Phi_{k_f,l}^*(\bm r) \Phi_{1s}(\bm r) \approx a_B K_1 D(k_f),
\end{equation}
where
\begin{equation}
D(k_f) = {\mathrm i a_B\over 2} \int d\bm r \: r \cos{\varphi}\Phi_{k_f,1}^*(\bm r) \Phi_{1s}(r) \ll 1.
\end{equation}
Within the hydrogenic model, which is used hereafter for crude estimations, $D(k_f) \propto (k_fa_B)^{-3}$ and the ratio of the interband contribution~\eqref{dir:exc} and the intraband contribution~\eqref{M_direct_intra} can be estimated as
\begin{equation}
\left| {M_{dir} \over M_{dir}'} \right| \sim {\gamma_6 \over a_B  E_g'} \left({E_g\over E_B}\right)^{3/2} \sim  {E_g\over E_B} \gg 1.
\end{equation}
Thus, the resonant interband process is dominant. Similar estimate holds for the exchange contributions.

\subsection{Auger recombination rate}

The exciton Auger recombination rate at resonant interband scattering of two $1s$ excitons into the $ns$ exciton state is given by [cf. Eq.~\eqref{RA} of the main text]
\begin{multline}
	R_A n_{X}^2 = {2\pi \over \hbar} \sum_{\bm K_1, \bm K_2} |M_{exch}(\bm K_1, \bm K_2, \bm K_f,ns) |^2 f(K_1)f(K_2)\times \\
	\delta[E_g -2E_B - E_g' +E(K_1)+E(K_2)-E_{ns}(K_f)].
\end{multline}
In order to calculate the rate of transitions we take into account that the matrix element $M_{exch}(\bm K_1, \bm K_2, \bm K_f,ns)$ depends on the principal quantum number $n$ of the final state and is independent of the initial wavevectors of excitons.

Let us first assume that there is exact resonance, i.e., for the certain value of $n$ at $\bm K_1 = \bm K_2 = \bm K_f$ we have
\begin{equation}
\label{resonance:n}
E_g = E_g'+ 2E_B - E_{B,n},
\end{equation}
where $E_{B,n}$ is the binding energy of $ns$ state. Removing the energy conservation $\delta$-function and assuming that 
\begin{equation}
\label{Boltzmann}
f(K) = \mathcal N \exp{\left(- \frac{\hbar^2 K^2}{2M k_B T}\right)},
\end{equation}
i.e., the excitons are distributed according to the Boltzmann law at the temperature $T$, $\mathcal N$ is the normalization constant determined from the condition $$n_X=g\sum_{\bm K} f(K),$$
where the factor $g$ accounts for the spin and valley degeneracy, we have
\begin{equation}
	R_A = R_n, \quad R_n = 
	{\pi \over \hbar k_B T}\left| {2\pi e^2 A_n \: \gamma_3^* \gamma_6\over \varkappa a_B^2 E_gE_g'} \right|^2.
\end{equation}
Note, that the Auger process is active for collisions of bright (spin-allowed) excitons with bright or dark ones, while for the dark-dark scattering the process is strongly suppressed. In the latter case the $\bm k\cdot \bm p$ admixture with the valence band is minor and the recombination via discussed channel is not effective. The Auger decay rate can be recast in the alternative form
\begin{equation}	
R_A n_X^2	\equiv {n_X \over \tau_A},
\end{equation}
where we introduced the Auger recombination time $\tau_A(n_{X})$. Since $e^2/a_B \sim E_B$, we have an estimate in the case of the resonance with $1s$ state
\begin{equation}
	{1 \over \tau_A} 
	\sim 
	{n_X \over \hbar k_\text{B}T} \left(E_g {a_0^2 \over a_B} \right)^2,
\end{equation}
where $a_0$ is the lattice constant. At $T=4$~K, $a_0=3$~\AA, $a_B=1$~nm, $E_B=0.5$~eV, and the exciton density $n_X=10^9$~cm$^{-2}$ this estimate yields $\tau_A \sim 25$~ps.

Let us now take into account the detuning
\begin{small}
\begin{equation}
	\Delta = E_g' - E_g -  E_{B,n} + 2E_{B}.
\end{equation}
\end{small}

We have the sum over $\bm K_{1,2}$ in the following form:
\begin{small}
\begin{multline}
	\sum_{\bm K_1, \bm K_2} f(K_1)f(K_2) \delta\left[{\hbar^2(K_1^2+K_2^2-|\bm K_1+\bm K_2|^2)\over 2M} - \Delta \right] \\
= {n_X^2\over 2 k_\text{B}T} \text{e}^{-|\Delta|/k_\text{B}T}. 
\end{multline}
\end{small}
We see that the difference with the case of zero detuning is the exponent. The Auger recombination rate is given by
\begin{equation}
	R_A = R_n\text{e}^{-{|\Delta|}/k_\text{B}T}.
\end{equation}

\end{document}